\documentclass[12pt,reqno]{iopart}
\usepackage{color,amsfonts,amssymb,amsthm}
 
\newtheorem{theorem}{Theorem}
\def\Hil{\mathcal{H}}

\begin{document}
\title{Tensorial description of quantum mechanics}

\author{J. Clemente-Gallardo}
\address{ BIFI-Departamento de F\'{\i}sica Te\'orica and
Unidad Asociada IQFR-BIFI,  Universidad de
  Zaragoza,  Edificio I+D-Campus   R\'{\i}o Ebro, Mariano Esquillor
  s/n, 50018 Zaragoza (Spain)}  
\ead{jesus.clementegallardo@bifi.es}

\author{G. Marmo}
\address{
Dipartimento di Scienze Fisiche dell'Universit\`a di
 Napoli ``Federico II'' and INFN--Sezione di Napoli, Complesso
  Universitario di Monte Sant'Angelo, via Cintia, I-80126 Napoli,
  (Italy)}
\ead{marmo@na.infn.it}

\begin{abstract}

Relevant algebraic structures for the description of Quantum Mechanics
in the Heisenberg picture are replaced by tensorfields on the space of
states. 
This replacement introduces a differential geometric point of view
which allows for a covariant formulation of quantum mechanics under
the full diffeomorphism group.

\end{abstract}

\section{Introduction}
One important motivation for a geometrical formulation of Quantum
Mechanics originates from the following observation. The unification of fundamental
interactions requires that we unify, preliminarly, two distinct
theories: classical General Relativity and Quantum Mechanics.  It is
reasonable to expect that what will eventually emerge will entail a
substantial revision of both theories, which we may expect,  will
emerge merely as approximations to some underlying new 
theory which encompasses them. 

Other motivations for our geometrical formulation are the
following. The space of pure states of a quantum system (usually
identified with rays of a Hilbert space  $\mathcal{H}$ or with rank-one projectors
defined on $\mathcal{H}$) is not a linear space, rather it is a
Hilbert manifold (i.e., a differential manifold whose
tangent spaces (at each point) are endowed with a Hilbert space
structure). Additional instances of manifolds of states,
instead of a representation as elements of linear spaces, are provided
by coherent states and generalized coherent states
(see \cite{a1,a2}). Further, the space of entangled states, in composite
systems, does not carry a linear structure and therefore a manifold
point of view may help in studying their properties. 

The manifold point of view requires to consider nonlinear
transformations, say diffeomorphisms, which provide a point of view for
Quantum Mechanics closer to the one achieved in the transition from
Special to General Relativity. See \cite{4} for a detailed review of
the topic.

\section{The description of Quantum Mechanics}
\subsection{The carrier space of states and probability distributions}
The introduction of the geometrical point of view seems also more
close to what is being measured in the laboratory. This formulation
deals with expectation values of observables on the space of physical
states.  From this point of view, let us analyze the pairing between
physical observables and physical states. We recall that in the Schr\"odinger 
picture we associate a Hilbert space  with any quantum
system, physical observables being Hermitian operators acting on
it. It is important to notice that  the elements of the Hilbert space
do not correspond directly to states of our physical system,  instead,
they can be considered as convenient entities which can be used to define probability
distributions associated to  Hermitian operators.  

For instance, let us consider  a resolution of the
identity, associated with some self-adjoint operator.
It can be written as:
\begin{itemize}
\item $\sum_k |e_k\rangle\langle e_k|=\mathbb{I}_\mathcal{H}$ for $\{
  |e_k\rangle\}$ a discrete ortonormal basis of the Hilbert space
  $\mathcal{H}$, say eigenstates of an operator $E$, or
\item $\int |a\rangle da \langle a|=\mathbb{I}_\mathcal{H}$ for $\{
  |a\rangle\} $ an ortonormal basis of
  eigenstates of an operator $A$ which is supposed to have a
  continuous spectrum . 
\end{itemize}
The corresponding probability distribution is defined by:
\begin{equation}
  \label{eq:1}
  \mathcal{P}^E_\psi(j)=\frac{\langle
    \psi|e_j\rangle\langle e_j|\psi\rangle}{\langle\psi|\psi\rangle}, 
\end{equation}
for the case of the discrete basis and by
\begin{equation}
  \label{eq:2}
  d\mathcal{P}^A_\psi (a)=\frac{\langle
    \psi|a\rangle da\langle a|\psi\rangle}{\langle\psi|\psi\rangle},
\end{equation}
for the continuous case; in both cases for any Hilbert space vector
$|\psi\rangle \in \mathcal{H}$.

An important spin-off of the replacement of states with probability
distributions is that it calls for an extension of the notion of
state. Indeed, if we consider states to be  probability distributions,
a convex combination of them is  a probability distribution too
and therefore acceptable as a state of our physical system. Hence,
they may be used to define expectation values of the operators
representing physical observables. 

By using the identification of pure states with rank-one projectors on
the Hilbert space $\mathcal{H}$, as
$$
\mathcal{P}_{|\psi\rangle}= \frac{|\psi\rangle\langle\psi|}{\langle \psi|\psi\rangle},
$$
it follows that a convex combination
\begin{equation}
  \label{eq:3}
  \rho=\sum_k \lambda_k \frac{|\psi_k\rangle\langle\psi_k|}{\langle
    \psi_k|\psi_k\rangle}  \qquad \lambda_k\geq 0 \quad \sum_j \lambda_j=1
\end{equation}
is also an admissible state. We shall denote by $\mathcal{S}$ the set
of all possible states. 

Each point $\rho\in \mathcal{S}$ 
is associated with a convex combination of the corresponding
probability distributions:
\begin{equation}
  \label{eq:4}
  \sum_k \lambda_k \frac{\langle
    \psi_k|e_j\rangle\langle
    e_j|\psi_k\rangle}{\langle\psi_k|\psi_k\rangle} =\sum_k \lambda_k
  \mathcal{P}_{|\psi_k\rangle}^E(j)= \mathcal{P}_\rho^E(j).
\end{equation}

Thus our starting point will be to consider the space of states as
carrier space for the description of our quantum system and the space
of real valued functions on it as candidates for representing
physical observables.  Notice that in this picture, as we give up the
linear structure of the carrier space, there is no room for linear
transformations, linear operators or linear superposition of vectors
(see \cite{8,9}). 

\subsection{The algebraic structures of the space of physical observables}

In order to identify the geometrical (tensorial) structures with which
we should endow our carrier space, let us consider the structures
which emerge from it if we start from the usual
Hilbert space point of view.

The space of observables (i.e. of self-adjoint operators acting on
$\Hil$) corresponds (up to an imaginary factor) to
$\mathfrak{u}(\Hil)$, the real Lie algebra of the unitary group of
$U(\mathcal{H})$. It may
be identified with its dual $\mathfrak{u}^*(\Hil)$ by the (regular) scalar
product defined as 
$$
\langle\cdot|\cdot \rangle: \mathfrak{u}(\mathcal{H})\times
\mathfrak{u}(\mathcal{H})\to \mathbb{R}; \qquad \langle A | B\rangle
=\frac 12 \mathrm{Tr} AB, \quad \forall A, B\in \mathfrak{u}(\mathcal{H}).
$$ 
The corresponding isomorphism
\begin{equation}
  \label{eq:27}
 \zeta :\mathfrak{u}(\mathcal{H})\to \mathfrak{u}^*(\mathcal{H})
\end{equation}
 allows us to export the geometric
and algebraic structures existing in each space, into the other. We
can therefore consider the Poisson structure on the dual
$\mathfrak{u}^*(\mathcal{H})$ as a tensor field on the space of observables
(and therefore we can consider Hamiltonian dynamics),
or extend the Jordan structure defined on $\mathfrak{u}(\mathcal{H})$
into its dual.  We shall see this in the next section.
 
On the other hand, for arbitrary linear operators on the vector space
$\mathcal{H}$, we know that the associative product of two operators can
always be decomposed as the sum of a symmetric and a skew-symmetric
part:
\begin{equation}
AB=\frac 12 (AB+BA) +\frac 12 (AB-BA) \qquad \forall A, B\in
\mathrm{End}(\mathcal{H}). \label{eq:28}
\end{equation}
If we restrict the operation to the space of Hermitian operators, we
notice that the symmetric part is an inner operation, while the
skew-symmetric part is not.  This can be amended by introducing the
imaginary unit as a prefactor defining thus:
\begin{equation}
  \label{eq:9}
  A\circ B=AB+BA \qquad [A, B]=-i(AB-BA)
\end{equation}
Both structures are then inner in the space of Hermitian
operators. The skew-symmetric one defines a Lie-algebra structure, which
corresponds  to the unitary Lie algebra $\mathfrak{u}(\mathcal{H})$).
The symmetric operation defines the notion of Jordan algebra. This
structure was considered and completely analyzed in finite dimensions
in the famous paper by Jordan, von Neumann and Wigner \cite{6b}.

Jordan algebras are commutative but not associative. They satisfy a
weaker form of associativity, namely:
\begin{equation}
  \label{eq:10}
  (A\circ A)\circ (B\circ A)=((A\circ A)\circ B)\circ A \qquad \forall
  A, B \in \mathfrak{u}(\mathcal{H}). 
\end{equation}

Moreover, a compatibility condition between the two operations (the
Jordan and Lie algebra) holds true, namely
\begin{equation}
  \label{eq:11}
  [A, (B\circ C)]=[A, B]\circ C+B\circ [A, C].
\end{equation}
In the abstract setting, this additional relation together with the
two structures defines what is called a Lie-Jordan structure
(see \cite{1, 3,7}). 

The further requirement
\begin{equation}
  \label{eq:12}
  A\circ (B\circ C)-(A\circ B)\circ C=K[[A, C],B]
\end{equation}
where $K$ is a suitable real number, allows to reconstruct the associative product on the abstract space
with elements  $(A, B, C, \cdots)$.

\section{Tensorial description of the space of observables}

Our goal is to encode the algebraic structures existing for the set of
physical observables at the level of the expectation-value functions. 
First, let us associate with any operator $A:\mathcal{H}\to \mathcal{H}$, the
expectation value function $e_A:\mathcal{H} \to \mathbb{C}$ defined as
\begin{equation}
\label{eq:26}
  e_A(\psi)=\frac{\langle\psi|A\psi\rangle}{\langle\psi|\psi\rangle}.
\end{equation}
Any linear operator $A:\mathcal{H}\to \mathcal{H}$ can be decomposed
as the sum of its real and imaginary parts
$$
A=A_R+iA_I.
$$

Both,$A_R$  and $A_I$ are Hermitian operators. The expectation value
functions associated with Hermitian operators are real valued. 

\subsection{Poisson, Jordan  and associative structures}
We may introduce dynamics in this framework with the help of the Ehrenfest
picture and write the equations of motion in terms of the expectation
value functions:
\begin{equation}
  \label{eq:5}
  \frac {d}{dt} e_A(\psi)=\frac 1{i\hbar}\frac{\langle \psi|
    (HA-AH)\psi\rangle}{\langle\psi|\psi\rangle}=\frac{\langle \psi|
    [H,A]\psi\rangle}{\langle\psi|\psi\rangle},
\end{equation}
where 
\begin{equation}
  \label{eq:6}
  [H, A]:=\frac 1{i\hbar}(HA-AH).
\end{equation}
Equation (\ref{eq:5}) uses the Lie-algebraic structure defined by the
commutator on the space of Hermitian operators. 

Observables  also give
rise to observed quantities  with associated indetermination, i.e.,
the variance  
\begin{equation}
  \label{eq:8}
  \Delta A(\psi)=\frac{\langle \psi|
   A^2\psi\rangle}{\langle\psi|\psi\rangle}- \left ( \frac{\langle \psi|
    A\psi\rangle}{\langle\psi|\psi\rangle}\right )^2,
\end{equation}
and higher order moments. The Jordan structure of
the space of observables is reflected in this type of objects. 

Finally, we can also introduce the associative product of the linear
operators on $\mathcal{H}$ into the expectation-value functions.  By
using Eq. (\ref{eq:28}), it is immediate to write:
\begin{equation}
  \label{eq:29}
  e_A\star e_B(\psi):= e_{AB}(\psi)=\frac 12 e_{A\circ B}(\psi) + \frac {i\hbar}2 e_{[A, B]}(\psi)
\end{equation}

\subsection{Digression: algebraic structures described by tensor fields}
Given a vector space $V$ and a binary bilinear product
$$
\mathcal{B}:V\times V\to V,
$$
we can replace $V$ by its bidual, i.e., the set of linear functions on
the dual space

\begin{eqnarray*}
  V&\hookrightarrow \mathcal{F}(V^*); \qquad
  v&\mapsto \hat v.
\end{eqnarray*}

Then, we can define a tensor field  on $V^*$ which, at the
point $\xi\in V^*$ takes the form
\begin{equation}
  \label{eq:13}
  \tau_\mathcal{B}(d\hat v_1, d\hat v_2)(\xi)=\xi(\mathcal{B}(v_1,
  v_2)) \qquad \forall v_1, v_2\in V, \,\, \xi\in V^*,
\end{equation}
where $\hat v_1$ and $\hat v_2$ are the linear functions belonging to
$\mathcal{F}(V^*)$ which are defined by $v_1$ and $v_2$
respectively.

If we want a description
solely in terms of structures on the space of expectation value
functions, and using our experience from Classical Mechanics, we
should seek for Poisson and Jordan structures defined on that space ( we may
call them  \textit{Quantum Poisson (Quantum Jordan) structure(s)}), such that: 
\begin{equation}
  \label{eq:7}
  \{ e_A, e_B\} (\psi)=e_{[A, B]}(\psi); \qquad \{e_A,
  e_B\}_+(\psi)=e_{A\circ B}(\psi).
\end{equation}
As a result, the carrier space of states gets endowed with a
contravariant tensor $\Lambda$ which encodes the Poisson structure and
with a contravariant tensor $G$ which encodes the Jordan product.

In our case, the vector space $V$ represents the space of
expectation-value functions, which are functions on the carrier space
of the system. This carrier space may be identified with the  convex body
$\mathcal{S}$ defined by Equation (\ref{eq:3}), and therefore may be
considered to be
contained in $\mathfrak{u}^*(\mathcal{H})$. 
Denoting by $\rho\in \mathcal{S}$ a generic state, we can write the expectation value
function associated to a Hermitian operator $A$ as
\begin{equation}
  \label{eq:14}
  e_A(\rho)= \rho(A)=\mathrm{Tr}(\rho A),
\end{equation}
where we used the same symbol $\rho$ to represent the Hermitian
operator which is associated with the state $\rho\in
\mathfrak{u}^*(\mathcal{H})$ via the isomorphism $\zeta$ in Eq. (\ref{eq:27}).
This is a linear function on the dual space
$\mathfrak{u}^*(\mathcal{H})$, and thus it is associated (because of the
isomorphism  defined by Eq. (\ref{eq:27})) with the Hermitian
operator $A$. Thus,  the tensor
associated to  a bilinear binary structure
$\mathcal{B}:\mathfrak{u}(\mathcal{H})\times
\mathfrak{u}(\mathcal{H})\to \mathfrak{u}(\mathcal{H})$ where $(A,
B)\mapsto \mathcal{B}(A, B)$, reads
\begin{equation}
  \label{eq:15}
  \tau_\mathcal{B}(de_A, de_B)(\rho)=\rho(\mathcal{B}(A, B)).
\end{equation}

Hence we can now define the two tensors we are looking for:
\begin{itemize}
\item the Poisson tensor associated to the Lie-algebra structure that
  is defined on  $\mathfrak{u}^*(\mathcal{H})$ and reads:
  \begin{equation}
    \label{eq:16}
    \Lambda(de_A, de_B)(\rho)=\rho ([A, B]).
  \end{equation}
\item the Jordan tensor associated to the symmetric operation $\circ$
  defined on $\mathfrak{u}^*(\mathcal{H})$:
  \begin{equation}
    \label{eq:17}
    G(de_A, de_B)(\rho)=\rho (A\circ B).
  \end{equation}
\end{itemize}

Therefore we have been able to reconstruct the algebraic structures of
the space of physical observables as tensorial objects defined on the
space of expectation-value functions. Further details can be found in
\cite{5,6}.

\subsection{Characterizing the observables}
The question we want to answer now is the following: if we are given a
function $F$ in the space of states $\mathcal{S}$, how do we know that
it is associated with an observable?

We can answer the question by using the tensors $\Lambda$ and $G$ we
introduced in the previous section. Consider the Hamiltonian vector
field $X_F$ which is associated with $F$ via the Poisson tensor
$\Lambda$. This tensor field can be considered as the infinitesimal
generator of a diffeomorphism acting on $\mathcal{S}$. Then, it can be
proved (see \cite{2} ) that
\begin{theorem}
  $F$ represents a physical observable if and only if the associated
  Hamiltonian vector field $X_F$ preserves the tensor $G$, i.e.
  \begin{equation}
    \label{eq:18}
    \mathcal{L}_{X_F}G=0.
  \end{equation}
\end{theorem}

In this manner, we can affirm that the space of states $\mathcal{S}$
(with its structure as a stratified manifold (see \cite{5,6}),
along with the tensors $\Lambda$ and $G$, completely encodes Quantum
Mechanics at a geometric level. The description allows us to consider generic
transformations, not just the linear ones, since the emerging
description is covariant under the full diffeomorphism group. 

\section{Example: the two level system}
Let us illustrate our arguments in the simplest but nontrivial
example: the two level system. 

In this case, the set of Hermitian operators correspond (modulo a
multiplication by the imaginary unit) to the Lie algebra
$\mathfrak{u}(2)$. We can consider the basis defined by the four Pauli
matrices $\{ \sigma_\mu\}_{\mu=0, 1, 2, 3}$ and represent the
corresponding coordinates as:
\begin{equation}
  \label{eq:19}
  Y_\mu(A)=\frac 12 \mathrm{Tr}(A \sigma_\mu).
\end{equation}

The convex body of states $\mathcal{S}$ is defined by the set of
points 
\begin{equation}
  \label{eq:20}
  \rho=\frac 12( \sigma_0+\vec x \vec \sigma) \qquad \vec x \vec x\leq 1.
\end{equation}
Thus, physical states are in one-to-one correspondence with points
of the Bloch ball, the points in the surface ($\vec x\vec x=1$)
representing the pure states.

As for the tensor fields, it is simple to write them down by using
Eq (\ref{eq:15}):
\begin{equation}
  \label{eq:21}
  G=\frac{\partial}{\partial Y_0}\otimes Y_k\frac{\partial}{\partial
    Y_k} +Y_k\frac{\partial}{\partial
    Y_k}\otimes\frac{\partial}{\partial Y_0} +Y_0 \left ( 
\frac{\partial}{\partial Y_0}\otimes \frac{\partial}{\partial Y_0}+
\frac{\partial}{\partial Y_k}\otimes \frac{\partial}{\partial Y_k}\right )
\end{equation}
\begin{equation}
  \label{eq:22}
  \Lambda=\epsilon^{jkl}Y_j\frac{\partial}{\partial Y_k}\otimes\frac{\partial}{\partial
    Y_l} 
\end{equation}
By using this last tensor, we can, for instance, recover at the level
of $\mathfrak{u}^*(2)$, the angular-momentum commutation relations:
$$
\{ Y_k, Y_l\} =\epsilon_{klm}Y_m.
$$

By using the composition $\mathcal{A}=G+i\Lambda$, we can recover the associative
product of operators at the level of the expectation value
functions. Thus we can write:
\begin{equation}
  \label{eq:22}
  e_A\star e_B:= e_{AB}=\mathcal{A}_{\mu \nu}^\xi Y_\xi\frac{\partial
    e_A}{\partial Y_\mu}\frac{\partial
    e_B}{\partial Y_\nu}
\end{equation}

At this point, we can show how it is possible to use nonlinear
transformation on the carrier space. We can perform the transition to
polar spherical coordinates $(y_0, r, \theta, \phi)$
\begin{equation}
  \label{eq:23}
  x= r\sin \theta\sin \phi, \qquad y=r\sin \theta \cos \phi; \qquad 
z=r\cos \theta,
\end{equation}
and write the expression of the tensors in these coordinates. For
instance, the Poisson tensor becomes:
\begin{equation}
  \label{eq:24}
  \Lambda=\frac {1}{r\sin \theta}\frac{\partial}{\partial
    \theta}\wedge \frac{\partial}{\partial \phi},
\end{equation}
while the Jordan one reads:
\begin{eqnarray}
  \label{eq:25}
  G=&r\frac{\partial}{\partial y_0}\otimes \frac{\partial}{\partial
    r}+ r\frac{\partial}{\partial r}\otimes \frac{\partial}{\partial
    y_0}+\nonumber \\
&y_0 \left ( \frac{\partial}{\partial
    y_0}\otimes\frac{\partial}{\partial y_0}+ \frac{\partial}{\partial 
   r}\otimes\frac{\partial}{\partial r}+ \frac 1{r^2}\frac
 {\partial}{\partial 
    \theta}\otimes\frac{\partial}{\partial \theta}+ \frac
  1{r^2\sin^2\theta}\frac{\partial}{\partial 
    \phi}\otimes\frac{\partial}{\partial \phi}\right )
\end{eqnarray}

These tensorial expressions show that the equations of
motion, or the indetermination relations associated with pairs of
expectation value functions may be written  in any coordinate system.

\textbf{Acknowledgements}

G.M. would like to acknowledge the support provided by the Santander/UCIIIM
Chair of Excellence programme 2011-2012. J C-G has been partially supported by
DGA Grant (24/1) and MICINN Grant (Fis2009-12648-C03-02)

\end{document}